\documentclass[journal,comsoc]{IEEEtran}
\usepackage{amsfonts}
\usepackage{amssymb}
\usepackage{cite}
\usepackage[cmex10]{amsmath}
\usepackage{float}
\usepackage{color}
\usepackage{stfloats,fancyhdr}
\usepackage{amsmath}
\usepackage{algorithm}
\usepackage{algorithmic}
\usepackage{multirow}
\usepackage{changepage}
\usepackage[normalem]{ulem}
\usepackage{amsmath}

\newtheorem{remark}{Remark}

\IEEEoverridecommandlockouts

\ifCLASSINFOpdf
  \usepackage[pdftex]{graphicx}
  \DeclareGraphicsExtensions{.pdf,.jpeg,.png}
\else
  \usepackage[dvips]{graphicx}
  \DeclareGraphicsExtensions{.pdf}
\fi

\usepackage{subfigure}

\usepackage{fancybox,dashbox}

\begin{document}
%
\title{{Distributed Multi-Relay Selection in Accumulate-then-Forward Energy Harvesting Relay Networks}
\thanks{Y. Gu, H. Chen, Y. Li, Y.-C. Liang and B. Vucetic are with School of Electrical and Information Engineering, The University of Sydney, Sydney, NSW 2006, Australia (email: yifan.gu@sydney.edu.au, he.chen@sydney.edu.au, yonghui.li@sydney.edu.au, liangyc@ieee.org, branka.vucetic@sydney.edu.au).}
\thanks{The material in this paper was presented in part at the IEEE International Conference on Communications, Kuala Lumpur, Malaysia, in May 2016.}
}

\author{Yifan Gu, He Chen, Yonghui Li, Ying-Chang Liang
and Branka Vucetic}

\maketitle

\begin{abstract}
This paper investigates a wireless-powered cooperative network (WPCN) consisting of one source-destination pair and multiple decode-and-forward (DF) relays. We develop an energy threshold based multi-relay selection (ETMRS) scheme for the considered WPCN. 
The proposed ETMRS scheme can be implemented in a fully distributed manner as the relays only need local information to switch between energy harvesting and information forwarding modes. By modeling the charging/discharging behaviours of the finite-capacity battery at each relay as a finite-state Markov Chain (MC), we derive an analytical expression for the system outage probability of the proposed ETMRS scheme over mixed Nakagami-$m$ and Rayleigh fading channels. Based on the derived expression, the optimal energy thresholds for all the relays corresponding to the minimum system outage probability can be obtained via an exhaustive search. However, this approach becomes computationally prohibitive when the number of relays and the associated number of battery energy levels is large. To resolve this issue, we propose a heuristic approach to optimize the energy threshold for each relay. To gain some useful insights for practical relay design, we also derive the upper bound for system outage probability corresponding to the case that all relays are equipped with infinite-capacity batteries. Numerical results validate our theoretical analysis. It is shown that the proposed heuristic approach can achieve a near-optimal system performance and our ETMRS scheme outperforms the existing single-relay selection scheme and common energy threshold scheme.
\end{abstract}


\IEEEpeerreviewmaketitle
\section{Introduction}
The performance of many wireless communication networks in practice is largely confined by the energy constrained devices that require replenishment periodically. Recently, a novel radio-frequency (RF) energy transfer and harvesting technique has been proposed as a new viable and promising solution to prolong the lifetime of energy constrained wireless networks \cite{RF-SURVEY1}. RF energy transfer and harvesting enables wireless devices to harvest energy from RF signals broadcast by ambient/dedicated energy transmitters to charge their batteries \cite{dc1,dc2}. This technique has opened a new research paradigm, termed wireless-powered communication (WPC), which has become a hot research topic recently (see, e.g., \cite{RF-SURVEY2,RF-SURVEY3} and references therein).

The WPC technique has brought new research opportunities to cooperative communications, which have attracted an upsurge of interest during the past decade due to its various advantages \cite{Yonghui_book_2010}. 
In this paper, we refer to a cooperative communication network with wireless-powered\footnote{Throughout this paper, we use the terms ``wireless-powered" and ``energy harvesting" interchangeably.} relay(s) as a wireless-powered cooperative network (WPCN). In \cite{Nasir-single-relay}, Nasir \emph{et. al} first investigated a classical three-node WPCN consisting of one source-destination pair and one energy harvesting amplify-and-forward (AF) relay. Two practical relaying protocols, namely time switching-based
relaying and power splitting-based relaying, were proposed and analyzed in \cite{Nasir-single-relay}. Inspired by this seminal work, a plenty of works focusing on the design and/or analysis of WPCNs have published in open literature very recently (see \cite{Chen_ICC_2014,zhiguoding-single-relay2,Fang_SPL_2015_Distributed,Zhong_TCom_2014_Wireless,Zeng_WCL_2015_Full,Krikidis_Tcom_2014_A,
Zhou_SPL_2014_Joint,Zhu_Tcom_2015_Wireless,Xiong_JSAC_2015_Wireless,caijunzhong-single-relay2,
zhiguoding-multi-relay,hechen,Chen_TWC_2015_Dis,r1,r2} and references therein).

All aforementioned works on WPCNs assumed that the wireless-powered node(s) exhausts the harvested energy in the current time slot to perform information transmission/forwarding straight away. Equipping each wireless-powered node with an energy storage (e.g., a rechargeable battery) such that they can accumulate the harvested energy and then perform information tranmission/forwarding in an appropriate time slot can improve the system performance significantly. The energy accumulation process of the classical three-node WPCNs with a single wireless-powered relay was modeled and the resulting network performance was analyzed in \cite{row-stac,Nasir_Tcom_2015_Wire} for finite and infinite storage scenario, respectively. \cite{review2} studied a multi-user network where all the users can harvest and accumulate energy from the base station simultaneously. Based on the considered system, the authors derived the transmission probability, the signal-to-interference ratio coverage probability and the overall success probability. It is also of great importance to investigate the network setup with multiple wireless-powered relays. Specifically, the relay selection problem was studied in \cite{review3} for a time-division and full-duplex block structure. Inspired by the max-max relay selection strategy, a new relay selection scheme is proposed that the relay with the best source-to-relay link is selected to receive information and store it in its buffer while the relay with the best relay-to-destination link is selected to transmit information. The corresponding outage probability and throughput were then analyzed. Inspired by the opportunistic relaying (OR) originally proposed in \cite{OR}, a battery-aware relay selection (BARS) scheme was proposed and analyzed. In the BARS, the relays with accumulated energy exceeding a predetermined threshold will first form a subset, which need to feedback their channel state information (CSI) to the source. Then, the ``best" relay among the subset with maximum end-to-end signal-to-noise ratio (SNR) is selected by the source to forward its information, while other relays harvest energy in the first hop. \cite{five-multi-relay} studied a WPCN with multiple randomly distributed relays. A distributed beamforming (DB) scheme was proposed that all relay nodes which are fully charged at the beginning of the transmission block form a forwarding subset. Among this subset of relays, the relays that are able to decode the source's signal create a virtual multiple antenna array and transmit source's signal to destination coherently.

In this paper we develop an efficient energy threshold based multi-relay selection (ETMRS) scheme with energy accumulation capability at each relay for WPCNs. In the proposed ETMRS, each relay can flexibly switch between energy harvesting and information forwarding modes in each transmission block. It should be noted that the proposed scheme is purely distributed and only the local battery status and local channel state information\footnote{For more information about CSI acquisition for energy harvesting devices, interested readers could see \cite{CSI1} and \cite{CSI2}.} (CSI) are required at each relay to perform mode selection, thus not involving extensive inter-relay information exchanges as in existing schemes. We also consider a practical and more general channel fading than \cite{or-eaccumu,five-multi-relay}, where source-relay links and relay-destination links are assumed to experience \emph{independent but not necessarily identical distributed} (i.n.i.d.) channel fading. Different from \cite{five-multi-relay}, we adopt a multiple-level battery model to characterize the charging/discharging behavior at relay batteries. In this case, the amount of harvested energy available at relays could be different. As such, to adequately exploit the relays, in our ETMRS scheme we set an individual energy threshold for each relay. This is in contrast to the existing schemes that considered independent and identical distributed channel fading and adopted a common energy threshold for all relays \cite{or-eaccumu,five-multi-relay}. The performance of the proposed ETMRS scheme can be further improved by jointly optimizing the energy thresholds of all relays. Moreover, 
compared with the DB scheme in which the weights in distributed beamforming for the relays within the forwarding set are obtained based on full CSI of the whole forwarding subset, the weight at each relay in the proposed ETMRS approach is calculated based only on the local CSI such that it can be implemented in a fully distributed manner. This can effectively reduce the network overhead and latency. Note that in our ERMRS scheme the relays that are selected to forward source's information to destination in each transmission block can be any combination of all relays, which depends on not only the instantaneous CSI but also the long-term evolution of all relay batteries. This makes the performance analysis and system design of the considered network a non-trivial task since we need to develop a systematic approach to characterize the probability of different relay combinations as well as the statistics of the corresponding end-to-end SNR as a sum of i.n.i.d random variables.  Furthermore, the energy threshold of each relay that determines its long-term energy evolution should be jointly designed to boost the system performance.

\textbf{\emph{Notation}}: Throughout this paper, we use $f_{X}(x)$ and $F_{X}(x)$ to denote the probability density function (PDF) and cumulative distribution function (CDF) of a random variable $X$. $\Gamma \left( {\cdot} \right)$ is the Gamma function \cite[Eq. (8.310)]{Tableofintegral}, $\gamma \left( {\cdot,\cdot} \right)$ is the lower incomplete gamma function \cite[Eq. (8.350.1)]{Tableofintegral} and $\left\lceil  \cdot  \right\rceil $ is the ceiling function. We use $\left( \cdot \right)^*$ and $\left( \cdot \right)^T$ to represent the complex conjugate and the transpose of a matrix or vector, respectively. $\mathbb{E}\left[ \cdot  \right]$ is the expectation operator and ${\bf I}$ denotes the identity matrix. $P\left\{ {A\left| B \right.} \right\}$ is the conditional probability of $A$ under a given condition $B$.

\section{System Model and Scheme Design}
\subsection{System Model}
\begin{figure}
\centering \scalebox{0.30}{\includegraphics{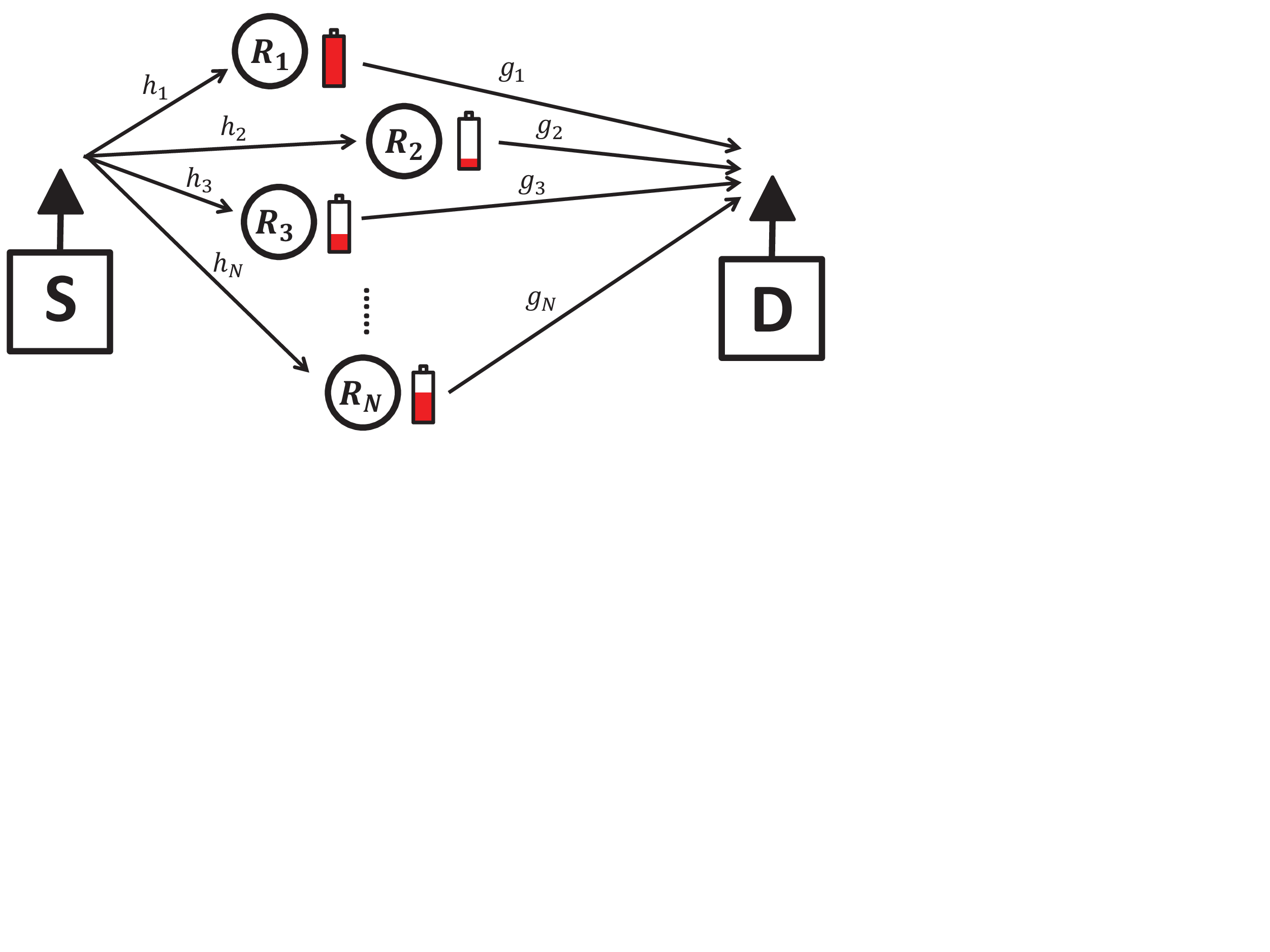}}
\caption{The considered WPCN with one source-destination pair and $N$ wireless-powered relays. }\label{fig:stm}
\end{figure}
As depicted in Fig. \ref{fig:stm}, in this paper we investigate a WPCN consisting of one source-destination pair and $N$ decode-and-forward (DF) relays, which are deployed to assist the source's information transmission. We assume that there is no direct link between source and destination due to obstacles or severe attenuation. Also, all nodes are equipped with single antenna and work in half-duplex mode. As in \cite{five-multi-relay}, we consider the scenario that all DF relays are wireless-powered devices and purely rely on the energy harvested from RF signals broadcast by source to perform information forwarding. Moreover, these relays are equipped with separate energy and information receivers \cite{time-switching}. As such, they can flexibly switch between energy harvesting (EH) mode and information forwarding (IF) mode at the beginning of each transmission block. We also assume that each relay is equipped with a finite-capacity rechargeable battery such that it can perform energy accumulation and scheduling across different transmission blocks. Specifically, they can accumulate the harvested energy to a certain amount before assisting the source's information transmission.

We denote by $T$ the duration of each transmission block, which is further divided into two time slots with equal length $T/2$. During the first time slot, the source broadcasts its signal to all relays. At each relay operating in EH mode, the received signal is delivered to the energy receiver to convert to direct current and charge the battery. In contrast, the received signal at each relay in IF mode is connected to its information receiver to decode the information sent by the source. All relays that operate in IF mode and decode the source's information correclty form a \emph{decoding set}. In the second time slot, all relays in the decoding set will jointly forward the source's information to destination by consuming part of the accumulated energy from their batteries. On the other hand, other relays outside the decoding set keep in silence during the second time slot\footnote{For simplicity, we consider that the relays out of the decoding set will not harvest energy in the second time slot since the amount of energy harvested from the forwarded signals would be negligible compared with the energy harvested from the source. }.

We hereafter use subscript-$S$ and subscript-$D$ to denote the source and destination respectively. We denote by ${R_u}$, $u \in \left\{ {1,2, \cdots ,N} \right\}$, the $u$-th wireless-powered relay. Among existing fading models, Rician fading would be the most appropriate one to characterize the channel fading of $S$-$R_u$ links. This is mainly motivated by the fact that the up-to-date wireless energy transfer techniques could only be operated within a relatively short communication range such that the line-of-sight (LoS) path is very likely to exist in these links. However, the statistical functions (e.g., cumulative density function (CDF) and probability density function (PDF)) of Rician fading are very complicated, which would make the analysis extremely difficult \cite{Zhong_Tcom_2015_Wireless}. Fortunately, the Rician distribution could be well approximated by the more tractable Nakagami-$m$ fading model. Thus, in this paper we adopt an asymmetric scenario for the fading distributions of source-to-relay links and relay-to-destination links. Specifically, the $S$-$R_u$ link is assumed to be subject to Nakagami-$m$ fading with fading severity parameter $m_u$ and average power gain ${\lambda _{SR_u}}$, while the $R_u$-$D$ link suffers from Rayleigh fading with average power gain ${\lambda _{R_u D}}$ as the distance between them may be much further. Besides, all channels in the considered system experience slow, independent, and frequency-flat fading such that instantaneous channel gains remain unchanged within each transmission block but change independently from one block to the other. It is worth mentioning that we do not require the channel fading parameters to be identical or non-identical for both hops. That is, we investigate a general \emph{independent but not necessarily identical} fading model, which includes the independent and identical one as well as the independent and non-identical one for special cases. Without loss of generality, we consider a normalized transmission block (i.e., $T=1$) hereafter.

Let $P$ denote the source transmit power and $x$ denote the transmitted symbol with $\mathbb{E} \left[{{\left| {{x}} \right|^2}}\right]=1$. The received signal at the $u$-th relay during the first time slot is thus given by
\begin{equation}\label{11}
y_u = \sqrt{P} h_u x + n,
\end{equation}
where $h_u$ is the channel coefficient between $S$ and $R_u$, and $n$ denotes the additive white Gaussian noise (AWGN) with zero mean and variance $N_0$ at the receiver side. 

When the $u$-th relay works in EH mode, the received signal $y_u$ will be delivered to its energy receiver and converted to direct current to charge the battery. The amount of harvested energy at $R_u$ during the first time slot can thus be expressed as
\begin{equation}\label{EH}
\tilde {E_{{u}}} =  {1 \over 2}\eta P{H_{u}},
\end{equation}
where $0 < \eta  < 1$ is the energy conversion efficiency and ${H_{u}}=  \left| {h_u } \right|^2$ is the channel power gain between $S$ and $R_u$. Note that in (\ref{EH}), we ignore the amount of energy harvested from the noise since the noise power is normally very small and below the sensitivity of the energy receiver. On the other hand, if $R_u$ opts to decode information in the first time slot, it will harvest zero energy.

Let $\Phi$ denote the current decoding set. In the second time slot, all relays in the decoding set $\Phi$ will jointly forward the source's information to the destination by implementing the distributed beamforming technique \cite{weightmaxsnr}. Specifically, the transmitted signal at $R_u \in \Phi$ is given~by
\begin{equation}\label{3}
x_u = {w_u}\sqrt {P_u} x,
\end{equation}
where ${w_u}$ is the weight of $R_u$ in distributed beamforming and $P_u$ is the transmit power of $R_u$. In the considered WPCN, we assume that each relay only knows its local CSI of the second hop. In this case, the optimal weight for the $u$-th relay that maximizes the overall end-to-end SNR can be expressed as $w _u  = {{g_{u}^* } / {\left| {g_{u} } \right|}}$ \cite{weightmaxsnr}, where ${g_u}$ is the complex channel coefficient between $R_u$-$D$. We define $\hat {g_u}=\left| {g_{{u}}} \right|$ for notation simplicity. The received signal at the destination can thus be expressed as
\begin{equation}
y_d = \sum\limits_{u:R_u \in \Phi } \hat {{{g_u}}} \sqrt {{{{{P_u} }}}} x + n.
\end{equation}
As a result, the conditional end-to-end SNR for a given decoding set $\Phi$ can be written as
\begin{equation}\label{1}
\gamma_\Phi   = {{{{\left( {\sum\limits_{u:R_u \in \Phi } \sqrt {{{{  {P_u} }}}}\hat {{{g_u}}} } \right)}^2}} \over {{N_0}}}.
\end{equation}

\subsection{Energy Threshold Based Multi-Relay Selection}

In this paper we develop an energy threshold based multi-relay selection (ETMRS) framework for the considered WPCN. In our ETMRS scheme, each relay $R_u$ determines its individual energy threshold, denoted by $\tilde \chi_u $. This energy threshold includes two parts: the first part is the energy consumption for circuit operation (e.g., information decoding), denoted by $\tilde \alpha$ for all the relays; the second part is the energy consumption for information forwarding, denoted by $ \tilde \beta_u$ for relay $R_u$. Each relay decides to operate in EH mode or IF mode based on its own battery status at the beginning of each transmission block. Specifically, relay $R_u$ will perform the IF operation only when its accumulated energy is not less than its associated energy threshold $\tilde {\chi_u }$. Otherwise, it will opt EH mode to further accumulate energy in its battery. Moreover, if $R_u$ works in the IF mode and falls in the decoding set, it will decode and forward the source signal to destination by consuming the amount of energy $\tilde {\beta_u }$ from its battery in the second time slot. The conditional end-to-end SNR given in (\ref{1}) can now be updated by substituting $P_u = \tilde {\beta_u} / (1/2)  = 2\tilde  {\beta_u}$. That is,
\begin{equation}\label{snr}
\gamma_\Phi   = {{{{\left( {\sum\limits_{u:R_u \in \Phi } \sqrt {{{{  {2 \tilde {\beta_u} }}}} } \hat {{{g_u}}}}\right)}^2}} \over {{N_0}}}.
\end{equation}


Note that we are investigating a distributed scheme such that each relay only requires local CSI and battery status to determine its operation modes. In this case, the optimal transmit power for each transmission block cannot be obtained as it requires global CSI and battery status. As a result, the fixed transmit power strategy is still preferable in our considered case and we can thus set fixed energy thresholds to each relay, $\tilde \chi_u, u=1,2,\cdots,N$. Moreover, the energy thresholds for all the relays should be different in order to achieve the best system performance when the relays are in different locations. For the special case that all the relays are co-located in a cluster, the channels of each hop are now independent and identically distributed (i.i.d.). In this sense, all relays can adopt the same energy threshold $\tilde {\chi} =  \tilde \chi_1 = \tilde \chi_2 =  \cdots  = \tilde \chi_N$ and consume the same power to forward information, denoted by $\tilde {\beta} =  \tilde \beta_1 = \tilde \beta_2 =  \cdots  = \tilde \beta_N$. We then have the conditional end-to-end SNR of this i.i.d. case given by
\begin{equation}\label{snr1}
\mathord{\buildrel{\lower3pt\hbox{$\scriptscriptstyle\frown$}}\over
 \gamma_\Phi  }  = {2 \tilde {\beta}{{{\left( {\sum\limits_{u:R_u \in \Phi } \hat {{{g_u}}} } \right)}^2}} \over {{N_0}}}.
\end{equation}

\section{Performance Analaysis}\label{sec:MC}
To analyze the performance of the proposed ETMRS scheme, in this section we first characterize the dynamic charging/discharging behaviors of the relay batteries. We consider a discrete-level and finite-capacity battery model. Thus, it is natural to use a finite-state Markov chain (MC) to model the dynamic behaviors of relays' batteries. From the MC model and the derived stationary distribution of the battery, we then derive an approximate analytical expression of the system outage probability for our ETMRS scheme. Based on the derived analytical expression, we subsequently discuss how to optimize the energy threshold of the relays to reduce the system outage probability.

\subsection{Markov Chain of Relay Batteries}
Thanks to the fact that our ETMRS scheme is a decentralized relay selection approach and the relays make decisions of their operation modes based only on their local CSI and battery status, we thus can evaluate the steady state distributions of all relays' batteries separately as they are independent to each other. In \cite{row-stac}, the authors investigated a single relay WPCN and proposed an adaptive information forwarding scheme such that the relay forwards the source information only when its residual energy can guarantee an outage-free transmission in the second hop. The transition probabilities of the relay battery was summarized into eight general cases. Recall that our ETMRS scheme implements fixed transmit powers for multiple relays. By adequately using this feature, a compact mode-based method is used to evaluate the transition probabilities of the MC for each relay.

Let $C$ denote the capacity of all relays' batteries and $L$ denote the number of discrete levels excluding the empty level in each battery. Then, the $i$-th energy level of the relay battery can be expressed as ${\varepsilon_i} = iC/L$, $i  \in \left\{  0,1,2 \cdots L\right\}$. It is worth pointing out that as shown in \cite{Huang_TWC_2008_Wire}, the adopted discrete battery model can tightly approximate its continuous counterpart when the number of energy levels (i.e., $L$) is sufficiently large, which will also be verified later in the simulation section. For each relay node, we define state ${S_i}$ as the relay residual energy in the battery being ${\varepsilon_i}$. The transition probability $T_{u}^{i,j}$ is defined as the probability of transition from state $S_i$ to state $S_j$ at the $u$-th relay. With the adopted discrete-level battery model, the amount of harvested energy can only be one of the discrete energy level. Thus, the discretized amount of harvested energy at the $u$-th relay during an EH operation is defined as
\begin{equation}\label{deh}
{E_u} \buildrel \Delta \over = {\varepsilon _{j}}, \quad {j} = \arg \mathop {\max }\limits_{i \in \left\{ {0,1, \cdots ,L} \right\}} \bigg\{{\varepsilon _i}:{\varepsilon _i} \le \tilde {E_{u}}\bigg\}.
\end{equation}
Moreover, for relays operating in the IF operation mode, the energy consumption for decoding operation $\alpha$ should also be discretized to one specific energy level of the battery with the definition give by
\begin{equation}\label{dif}
{\alpha} \buildrel \Delta \over = {\varepsilon _{j}}, \quad j=
\left\{
\begin{matrix}
\begin{split}
&\arg \mathop {\min }\limits_{i \in \left\{ {0,1, \cdots ,L} \right\}} \bigg\{{\varepsilon _i}:{\varepsilon _i} \ge \alpha \bigg\}, \text{if} \quad \tilde \alpha \le {\varepsilon _L} \\
&\infty, \text{if} \quad \tilde \alpha > {\varepsilon _L}
\end{split}
\end{matrix}
\right. \\
\end{equation}

The energy consumption for information forwarding $\tilde \beta_u$ and the energy threshold for each relay $\tilde \chi_u$ should be chosen from one of the energy levels of the battery excluding the empty level. The descretized energy consumption for information forwarding ${\beta_u}$ and discretized energy threshold for each relay  ${\chi_u}$ can be defined as $ {\beta_u} \in \left\{\varepsilon _{1}, \varepsilon _{2},\cdots, \varepsilon _{L} -\alpha \right\}$ and $\chi_u={\alpha}+{\beta_u} \in \left\{\alpha+\varepsilon _{1}, \alpha+ \varepsilon _{2},\cdots, \varepsilon _{L} \right\}$, respectively. Note that the system will work properly only when $\alpha < \varepsilon _{L}$, otherwise when $\alpha \ge \varepsilon _{L}$, the fully charged battery even cannot support the circuit energy consumption of the relay and $\beta_u$ does not exist.

We now evaluate the state transition probabilities of the MC for each relay $R_u$, $u \in \left\{ {1,2, \cdots ,N} \right\}$. Different from \cite{row-stac} that summarizes all the transition probabilities into eight general cases, we propose a compact mode-based approch which summarizes all the transition probabilities into the following two cases.
\subsubsection{The relay $R_u$ operates in EH mode (${S_i}\quad to\quad {S_j}$ with $  0 \le i  <{\chi_u \over \varepsilon _1} \le L$ and $\forall j$)}
When the relay $R_u$ operates in the EH mode, it harvests energy from the source during the first time slot while remains in silence during the second time slot. Due to the fact that the relay battery is not discharged in the transition, the transition probability is none-zero only when the end state falls into the set ${S_j} \in \left\{{S_i}, {S_{i+1}},\cdots, {S_L}\right\}$. Specifically, ${S_j} = {S_i}$ indicates that the battery level of the relay remains unchanged and the harvested energy during the transition $\tilde {E_{u}}$ is discretized to zero (i.e., ${E_u}=0$). ${S_j} = {S_L}$ denotes the case that the battery is fully charged during the transition and the harvested energy should be larger than $\varepsilon _{L-i}$. From the definition of discretization given in (\ref{deh}), the transition probabilities for $R_u$ operates in EH mode can be summarized in (\ref{10}) on top of the next page.

\begin{figure*}[!t]
\begin{equation}\label{10}
\begin{split}
{T_{u}^{i,j}}
  & = \left\{ {
\begin{matrix}
\begin{split}
   &{\Pr \left\{ \varepsilon _{j-i} \le \tilde {E_{u}} < \varepsilon _{j-i+1} \right\}, \quad \text{if} \quad i \le  j<L}  \\
   &{\Pr \left\{ \tilde {E_{u}} \ge \varepsilon _{L-i} \right\}, \quad \text{if} \quad i \le  j=L}  \\
   &{0, \quad \text{if} \quad i >  j }\\
\end{split}
\end{matrix}
  } \right. = \left\{ {
\begin{matrix}
\begin{split}
   &{\Pr \left\{ {{2\left({j-i}\right)C} \over \eta P L} \le H_u < {{2\left({j-i+1}\right)C} \over \eta P L} \right\}, \quad \text{if} \quad i \le  j<L}  \\
   &{\Pr \left\{ H_u \ge {{2\left({L-i}\right)C} \over \eta P L} \right\}, \quad \text{if} \quad i \le  j=L}  \\
   &{0, \quad \text{if} \quad i >  j }\\
\end{split}
\end{matrix}
  } \right..\\
\end{split}
\end{equation}
\hrulefill
\vspace*{4pt}
\end{figure*}

Recall that the channels between source and relays are assumed to suffer from Nakagami-$m$ fading. As such, the PDF and CDF of $H_u$ are given by \cite[eq.2.21]{fadingchannels}
$f_{H_{u}}(x) = {{{{b_{u}} ^{m_{u}}}} \over {\Gamma (m_{u})}}{x^{{m_{u}} - 1}}\exp ( - {b_{u}} x)$, ${F_{H_{u}}}\left( x \right) = {{\gamma \left( {m_{u},b_{u} x} \right)} \over {\Gamma \left( m_{u} \right)}}$, where $b_{u}  = m_u / {\lambda_{S R_u} }$. With the CDF, the transition probabilities for this case can now be expressed as

\begin{equation}
\begin{split}
{T_{u}^{i,j}}
    & = \left\{ {
\begin{matrix}
\begin{split}
   &{\begin{split}
   &{F_{{H_u}}}\left( {{{2(j-i+1)C} \over {\eta PL}}} \right)\\&-{F_{{H_u}}}\left( {{{2(j-i)C} \over {\eta PL}}} \right)
   \end{split}, \quad \text{if} \quad i \le  j<L}  \\
   &{1-{F_{{H_u}}}\left( {{{2(L-i)C} \over {\eta PL}}} \right), \quad \text{if} \quad i \le  j=L}  \\
   &{0, \quad \text{if} \quad i >  j }\\
\end{split}
\end{matrix}
  } \right..\\
\end{split}
\end{equation}

\subsubsection{The relay $R_u$ operates in IF mode (${S_i}\quad to\quad {S_j}$ with $ {\chi_u \over \varepsilon _1} \le i  \le  L$ and $\forall j$)}
In this case, the relay $R_u$ will try to decode the received signal and forward it to destination if the decoding is successful. In the first time slot, the relay consumes energy $\alpha$ to decode the received signal from $S$. If the decoding is unsuccessful, it remains in silence during the second time slot. On the other hand, if the relay $R_u$ decodes the information successfully, it forwards the decoded information to the destination by further consuming $\beta_u$ from its battery during the second time slot. We can now conclude that after the transition, the end state is none-zero only when $j = i- {\alpha \over \varepsilon _1}$ or $j = i- {\chi_u \over \varepsilon _1}$. Recall that $\varphi_u$ is the decoding indicator of relay $R_u$ and the transition probabilities for $R_u$ operating in IF mode can be summarized as
\begin{equation}
\begin{split}
{T_{u}^{i,j}}
    & = \left\{ {
\begin{matrix}
\begin{split}
   &{\Pr \left\{ {{\varphi _u} = 0} \right\}, \quad \text{if} \quad j = i- {\alpha \over \varepsilon _1}}  \\
   &{1-\Pr \left\{ {{\varphi _u} = 0} \right\}, \quad \text{if} \quad i \le  j=i- {\chi_u \over \varepsilon _1}}  \\
   &{0, \quad \text{Otherwise}}\\
\end{split}
\end{matrix}
  } \right..\\
\end{split}
\end{equation}
We now analyze the term $\Pr \left\{ {{\varphi _u} = 0} \right\}$ for the $u$-th relay. Let $\gamma_u = {{P{H_u}} \over {{N_0}}}$ denote the received SNR at relay $R_u$. The channel capacity of the $S$-$R_u$ link is given by ${\Theta _u}{\rm{ = }}{1 \over 2}{\log _2}\left( {1 + {\gamma _u}} \right)$. According to the channel capacity, the term $\Pr \left\{ {{\varphi _u} = 0} \right\}$ can be evaluated as
\begin{equation}\label{general}
\Pr \left\{ {{\varphi _u} = 0} \right\} = \Pr \left\{ {{\Theta _u} <\kappa} \right\} = {F_{{H_u}}}\left( {{{vN_0} \over {P}}} \right),
\end{equation}
where $\kappa$ is the system transmission rate and $v=2^{2{\kappa}}-1$ is the SNR threshold for system outage.

Based on the above analysis, the transition probabilities can be re-written as
\begin{equation}
\begin{split}
{T_{u}^{i,j}}
    & = \left\{ {
\begin{matrix}
\begin{split}
   &{{F_{{H_u}}}\left( {{{vN_0} \over {P}}} \right), \quad \text{if} \quad j = i- {\alpha \over \varepsilon _1}}  \\
   &{1-{F_{{H_u}}}\left( {{{vN_0} \over {P}}} \right), \quad \text{if} \quad i \le  j=i- {\chi_u \over \varepsilon _1}}  \\
   &{0, \quad \text{Otherwise}}\\
\end{split}
\end{matrix}
  } \right..\\
\end{split}
\end{equation}

We now define $\mathbf{Z}_u = ({T_{u}^{i,j}})$ to denote the $(L+1) \times (L+1)$ state transition matrix of the MC for each relay $R_u$, $u \in \left\{ {1,2, \cdots ,N} \right\}$. By using similar method in \cite{row-stac}, we can easily verify that the MC transition matrix $\mathbf{Z}_u$ derived from the above MC model is irreducible and row stochastic. Thus for each relay $R_u$, there must exist a unique stationary distribution $\pmb{\pi}_u$ that satisfies the following equation
\begin{equation}\label{solve}
\pmb {\pi} _u  =\left( {{{\pi} _{u,0}},{{\pi} _{u,1}}, \cdots, {{\pi} _{u,L}}} \right)^{T} = \left({\mathbf{Z}_u}\right)^{T} \pmb {\pi} _u,
\end{equation}
where ${{\pi} _{u,i}}$, $i \in \left\{ {0,1, \cdots ,L} \right\}
$, is the $i$-th component of $\pmb {\pi}_u$ representing the stationary distribution of the $i$-th energy level at relay $R_u$. The battery stationary distribution of relay $R_u$ can be solved from (\ref{solve}) and expressed as \cite{row-stac}
\begin{equation}\label{app-energy-distri}
 \pmb {\pi} _u = {\left( { \left({\mathbf{Z}_u}\right)^{T} - \mathbf{I} + \mathbf{B} }\right)^{ - 1}}\mathbf{b},
\end{equation}
where ${\mathbf{B}_{i,j}}=1, \forall i, j$ and $\mathbf{b}={(1,1, \cdots ,1)^T}$.

Moreover, when it comes to the i.i.d. channel model, all relays are equipped with an identical energy threshold and they have the same transition matrix. The corresponding identical stationary distribution of all relays can be similarly obtained as (\ref{app-energy-distri}) and denoted by $\pmb {\pi} _1= \pmb {\pi} _2=  \cdots  =  \pmb {\pi} _N =  \pmb {\pi}=  \left( {{{\pi} _{0}},{{\pi} _{1}}, \cdots, {{\pi} _{L}}} \right) ^{T}$.

\subsection{System Outage Probability}
With the above derived stationary distribution of the relay batteries, we now characterize the system outage probability of the proposed ETMRS scheme. Let ${\rm O}$ denote the outage event of the considered system employing our ETMRS scheme. According to the full probability theory, we can express the system outage probability~as 
\begin{equation}\label{outage}
P_{\rm{out}} = \Pr \left\{ \rm O \right\} = \sum\limits_{\Phi  \in \Lambda } {\Pr \left\{ \Phi  \right\}} \Pr \left\{ {{\rm O}\left| \Phi  \right.} \right\},
\end{equation}
where $\Lambda  = \left\{ {R_1 ,R_2  \cdots ,R_N } \right\}$ denotes the set of all relays in the considered network and incorporates the decoding set $\Phi$ as its subset, $\Pr \left\{ \Phi  \right\}$ denotes the probability that the current decoding set is $\Phi$, and $\Pr \left\{ {{\rm O}\left| \Phi  \right.} \right\}$ denotes the probability that system outage occurs under the decoding set $\Phi$. In order to further expand (\ref{outage}), we define ${\Phi _{k,n}}$ to denote the $n$-th $k$-subset of $\Lambda$ (i.e., the $n$-th $k$-subset of $\Lambda$ contains exactly $k$ elements, $k = 1, 2, \cdots ,N$, $n = 1,2, \cdots ,\binom{N}{k}$). Then the outage probability of the ETMRS scheme can be further expanded as
\begin{equation}\label{outage1}
P_{\rm{out}} = \Pr \left\{ \emptyset \right\} + \sum\limits_{k = 1}^N {\sum\limits_{n = 1}^{\binom{N}{k}} {\Pr \left\{ {{\Phi _{k,n}}} \right\}\Pr \left\{ {{\rm O}\left| {{\Phi _{k,n}}} \right.} \right\}}},
\end{equation}
by realizing that the system outage probability equals to one when the decoding set is an empty set $\emptyset$. The empty decoding set can be caused by two kinds of event: One is that none of the relays working in IF mode decodes the received signal from source correctly. The other is that  all relays operate in EH mode and no relay performs IF. Based on the derived stationary discrete distribution of relay batteries given in (\ref{app-energy-distri}), we can calculate the first probability term in (\ref{outage1}) as follows
\begin{equation}\label{firstterm}
\begin{split}
\Pr \left\{ \emptyset   \right\}  &= \prod\limits_{u:R_u  \in \Lambda } {\Pr \bigg\{ {\left\{ {\zeta _u  = \zeta _I ,\varphi _u  = 0} \right\} \cup \zeta _u  = \zeta _E } \bigg\}}   \\
   &= \prod\limits_{u:R_u  \in \Lambda } { \left( {\Pr \left\{{\varphi_u = 0}\right\} \sum\limits_{i =  {\chi_u}/\varepsilon_1}^L {\pi _{u,i} }  + \sum\limits_{i = 0}^{ {\chi_u}/\varepsilon_1 }{\pi _{u,i} }}\right)}.  \\
\end{split}
\end{equation}
Similarly, the term $\Pr\left\{{\Phi_{k,n}}\right\}$ can be computed as (\ref{secondterm}) on top of the next page.

\begin{figure*}[!t]
\begin{equation}\label{secondterm}
\begin{split}
\Pr \left\{ {{\Phi _{k,n}}} \right\} &= \prod\limits_{u:R_u  \in \Phi_{k,n} } {\Pr \left\{ {\zeta _u  = \zeta _I ,\varphi _u  = 1} \right\}}  \prod\limits_{u:R_u  \notin \Phi_{k,n} } {\Pr \bigg\{ {\left\{ {\zeta _u  = \zeta _I ,\varphi _u  = 0} \right\} \cup \zeta _u  = \zeta _E } \bigg\}}   \\
  &  = \prod\limits_{u:R_u  \in \Phi_{k,n} } { \left[{\left( {1 - \Pr \left\{{\varphi_u = 0}\right\}} \right)\sum\limits_{i ={\chi_u}/\varepsilon_1 }^L {\pi _{u,i} } } \right]}\  \prod\limits_{u:R_u  \notin \Phi_{k,n}} { \left( {\Pr \left\{{\varphi_u = 0}\right\} \sum\limits_{i = { {\chi_u}}/\varepsilon_1}^L {\pi _{u,i} }  + \sum\limits_{i = 0}^{ {\chi_u}/\varepsilon_1 }{\pi _{u,i} }}\right)} .  \\
\end{split}
\end{equation}

\hrulefill
\vspace*{4pt}
\end{figure*}

To evaluate the third probability term in (\ref{outage1}), we first characterize the distribution of the conditional end-to-end SNR for a given decoding set. For notation simplicity, we use $\gamma_{k,n}$ to denote the received SNR at the destination when the decoding set is $\Phi_{n,k}$. Recall that the conditional end-to-end SNR for a certain decoding set is given in (\ref{snr}), which includes a weighted sum of Rayleigh random variables. However, to the best knowledge of authors, the exact distribution of a weighted sum of Rayleigh random variables does not exist in open literature. As a result, we cannot further characterize the exact distribution of $\gamma_{k,n}$. Fortunately, with the aid of a tight approximation for the CDF of a weighted sum of Rayleigh random variables derived in \cite{appro2}, the CDF of $\gamma_{k,n}$ can be approximated as a gamma distribution and expressed as
\begin{equation}\label{5}
{F_{\gamma_{k,n}}} \left({x}\right) \approx  {{\gamma \left( {k,{{{{{{N_0}} \over { 4\sum\limits_{u:{R_u} \in {\Phi _{k,n}}} {{\beta _u}{\sigma _u^2}} } }}} }x} \right)} \over {\Gamma \left( k \right)}},
\end{equation}
where $\sigma_u = \sqrt {{\lambda _{{R_u}D}}/2}$ is the scale parameter of the Rayleigh fading channel between $R_u$ and~$D$.

\begin{table}
\centering \scalebox{0.51}{\includegraphics{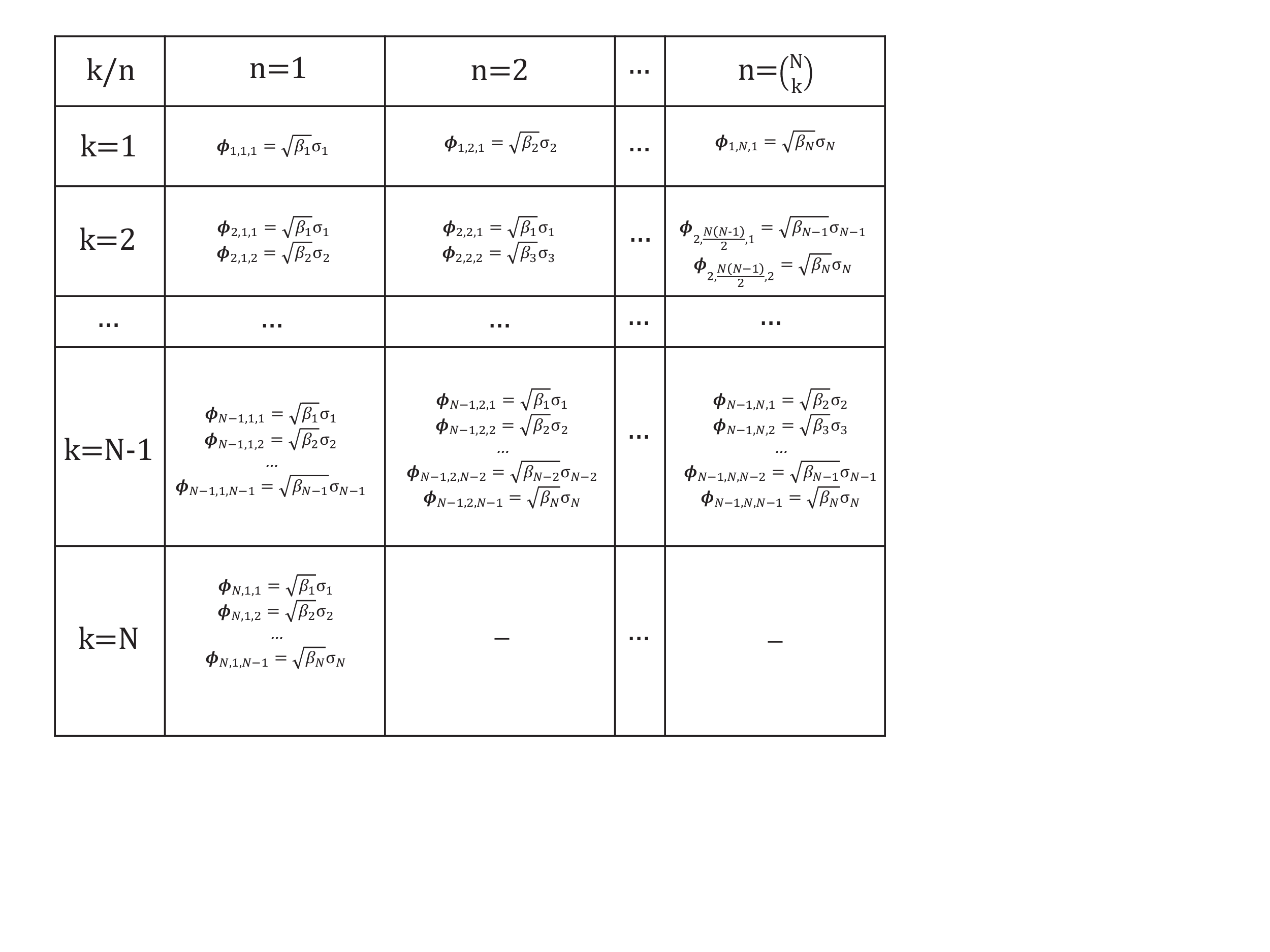}}
\caption{The relation between $\phi _{k,n,j}$ and $\sqrt{{\beta_u} }\sigma_u$, $u = 1,2, \cdots ,N$, for $k = 1, \cdots ,N$, $n = 1,2, \cdots ,\binom{N}{k} $ and $j = 1,2, \cdots ,k$. }\label{table:1}
\end{table}
In order to further expand the summation term in (\ref{5}), we use a similar method adopted in \cite{tablewtf,cl_chen,PRS}. To this end, we define a set ${\rm A}=\bigg\{ { \sqrt{ {\beta_u}}{\sigma_u}:{R_u} \in \Lambda ,u \in \left\{ {1,2, \cdots ,N} \right\}} \bigg\}$ with the same cardinality as the set $\Lambda$ (i.e., $\left| \rm A  \right| = \left| \Lambda  \right|$ ). Similarly, let ${\rm A}_{k,n}$ denote the $n$-th $k$-subset of $\rm A$. The $j$-th element of the subset ${\rm A}_{k,n}$ are denoted by $\phi _{k,n,j} \in {\rm A}_{k,n}$, $j = 1,2, \cdots ,k$. To be more clear, we list the corresponding relationship between $\phi _{k,n,j}$ and $\sqrt{{\beta_u} }\sigma_u$ in Table I.

With the corresponding relation between $\phi _{k,n,j}$ and $\sqrt{ {\beta_u} }\sigma_u$, (\ref{5}) can now be expressed as
\begin{equation}\label{condcdfsnr}
{F_{\gamma_{k,n}}} \left({x}\right) \approx  {{\gamma \left( {k,{a_{k,n}}x} \right)} \over {\Gamma \left( k \right)}}= 1 - \exp ( - a_{k,n} x) \sum\limits_{i = 0}^{k - 1} {{\left({a_{k,n} x}\right)^{i} } \over {i!}},
\end{equation}
where $a_{k,n}= {{{{{{N_0}} \over { 4 \sum\limits_{j = 1}^k {\phi_{k,n,j}^2}} }}} }$ and the last equality in (\ref{condcdfsnr}) holds according to \cite[Eq. (8.352.6)]{Tableofintegral} with integer $k$.
Based on the distribution of $\gamma_{k,n}$, the third probability term in (\ref{outage1}) can now be further expanded~as
\begin{equation}\label{2}
\begin{split}
\Pr \left\{ {{\rm O}\left| \Phi_{k,n}  \right.} \right\} &=  \Pr\left\{  {\gamma_{k,n}}< {v }\right\} ={F_{\gamma_{k,n}}}\left({v}\right) \\
&\approx 1 - \exp ( - a_{k,n} v) \sum\limits_{i = 0}^{k - 1} {{\left({a_{k,n} v}\right)^{i} } \over {i!}},
\end{split}
\end{equation}
where $v=2^{2{\kappa}}-1$ is the SNR threshold for system outage and $\kappa$ is the system transmission rate. By substituting (\ref{firstterm}), (\ref{secondterm}) and (\ref{2}) into (\ref{outage1}), we have derived an approximate analytical expression of the outage probability for the proposed ETMRS scheme.

In terms of the i.i.d channel fading case where the relays have the same energy threshold, the number of different decoding sets reduces to $N+1$ and the expression of system outage probability can be simplified to
\begin{equation}\label{iid1}
\mathord{\buildrel{\lower3pt\hbox{$\scriptscriptstyle\frown$}}\over
 P  }_{\rm{out}}= \Pr \left\{ \rm O \right\} = \Pr \left\{{ {\mathord{\buildrel{\lower3pt\hbox{$\scriptscriptstyle\frown$}}\over
 \emptyset  } }}\right\}  + \sum\limits_{k = 1}^N { {\Pr \left\{{ {{\mathord{\buildrel{\lower3pt\hbox{$\scriptscriptstyle\frown$}}\over\Phi} _{k}}} }\right\}\Pr \left\{{ {{\rm O}\left| {{\mathord{\buildrel{\lower3pt\hbox{$\scriptscriptstyle\frown$}}\over\Phi} _{k}}} \right.} }\right\}}},
\end{equation}
where $\Pr \left\{ {\mathord{\buildrel{\lower3pt\hbox{$\scriptscriptstyle\frown$}}\over
 \emptyset  } }\right\}$ is the probability that the decoding set is empty for the i.i.d. case, $\Pr \left\{ {{\mathord{\buildrel{\lower3pt\hbox{$\scriptscriptstyle\frown$}}\over\Phi} _{k}}} \right\}$ is the probability that the decoding set contains $k$ relays, and ${\Pr \left\{ {{\rm O}\left| {{\mathord{\buildrel{\lower3pt\hbox{$\scriptscriptstyle\frown$}}\over\Phi} _{k}}} \right.} \right\}}$ is the conditional outage probability when $k$ relays falls in the decoding set. Similar to the analysis of the general case, the first and second probability terms in (\ref{iid1}) can be expressed as
\begin{equation}\label{firsttermiid}
\begin{split}
\Pr \left\{ {\mathord{\buildrel{\lower3pt\hbox{$\scriptscriptstyle\frown$}}\over
 \emptyset  } }\right\} =  \left( {\Pr \left\{{\varphi = 0}\right\} \sum\limits_{i =  {\chi}/\varepsilon_1}^L {\pi _{i} }  + \sum\limits_{i = 0}^{ {\chi}/\varepsilon_1 }{\pi _{i} }}\right)^N, \\
\end{split}
\end{equation}
\begin{equation}\label{secondtermiid}
\begin{split}
\Pr \left\{ {{\mathord{\buildrel{\lower3pt\hbox{$\scriptscriptstyle\frown$}}\over\Phi }_{k}}} \right\} &=   {\binom{N}{k}} \left[{\left( {1 - \Pr \left\{{\varphi = 0}\right\} } \right)\sum\limits_{i ={\chi}/\varepsilon_1 }^L {\pi _{i} } } \right]^k \times\\
 &\quad \left( {\Pr \left\{{\varphi = 0}\right\} \sum\limits_{i = { {\chi}}/\varepsilon_1}^L {\pi _{i} }  + \sum\limits_{i = 0}^{ {\chi}/\varepsilon_1 }{\pi _{i} }}\right)^{N-k}, \\
\end{split}
\end{equation}
where $\chi = \alpha+\beta$ is the identical energy threshold for all the relays and $\Pr \left\{{\varphi = 0}\right\}$ is the outage probability of the relays for the special i.i.d. channel fading case, which can be derived from that for the general case given in (\ref{general}) and written as $\Pr \left\{{\varphi = 0}\right\} = {{\gamma \left( {m,  {m v N_0 \over {\lambda_{SR} P}}} \right)} \over {\Gamma \left( m \right)}}$, where $m$ is the identical severity parameter of the Nakagami-$m$ fading channels between source and relays and ${\lambda_{SR}}$ is the identical channel power gain of the first hop.

The third probability term in (\ref{iid1}) can be deduced from (\ref{2}) as
\begin{equation}\label{snriid}
\Pr \left\{{ {{\rm O}\left| {{\mathord{\buildrel{\lower3pt\hbox{$\scriptscriptstyle\frown$}}\over\Phi} _{k}}} \right.} }\right\} \approx 1 - \exp ( - {{{{{{N_0v}} \over { 4 k \beta \sigma^2} }}} }) \sum\limits_{i = 0}^{k - 1} {{\left({{{{{{{N_0v}} \over { 4 k \beta \sigma^2} }}} }}\right)^{i} } \over {i!}},
\end{equation}
where $\sigma= \sqrt {\lambda_{RD} /2} $ is the identical scale parameter for the Rayleigh fading channels between relays and destination. Note that different from (\ref{2}), the expression of the conditional system outage probability given in (\ref{snriid}) does not require the parameters defined in Table I. The outage probability for the special i.i.d. channel fading case can be obtained by substituting (\ref{firsttermiid}), (\ref{secondtermiid}) and (\ref{snriid}) into (\ref{iid1}).

\subsection{Energy Threshold Optimization}
We first consider the special i.i.d. channel fading case where the associated energy thresholds for all the relays is identical, denoted by $\chi = \alpha+\beta$. When the energy used for information forwarding $\beta$ grows, the conditional outage probability derived in (\ref{snriid}) reduces and the overall system outage probability correspondingly decreases. On the other hand, increasing of $\beta$ will lead to an increase of $\chi$. According to (\ref{secondtermiid}), it decreases the probability of each relay working in IF mode and the number of relays falling into the decoding set. This will increase the overall system outage probability. In simple words, when the designed energy threshold $\chi$ for the relays is small, most of relays could fall into the decoding set but their transmit power is low. When the energy threshold is large, only a few relays fall into the decoding set but their associated transmit power is high. As $\chi = \alpha+\beta \ge \alpha+ \varepsilon _{1}$, we can now infer that for the special i.i.d. channel fading case, there should exist an optimal energy threshold $\chi \in \left\{\alpha+\varepsilon _{1}, \alpha+\varepsilon _{2},\cdots, \varepsilon _{L} \right\}$ that minimize the system outage probability of the proposed ETMRS scheme. Similarly, for the general i.n.i.d. channel fading case where the relays may be located dispersively. There should also exist an optimal energy threshold set $\left\{\chi_1, \chi_2,\cdots,\chi_N\right\}, \chi_u \in \left\{\alpha+\varepsilon _{1}, \alpha+\varepsilon _{2},\cdots, \varepsilon _{L} \right\}, \forall u=1,2,\cdots,N$, for all relays that minimize the overall system outage probability.

Due to the complexity of the adopted MC model, it is difficult to derive an analytical expression for the optimal energy threshold of each relay. However, for the special i.i.d. channel fading case, the optimal energy threshold can be easily achieved by performing a one-dimensional exhaustive search from all the possible energy levels with the derived analytical outage probability given in (\ref{iid1}). The computation complexity of this search is given by $O\left(L\right)$.

When it comes to the general i.n.i.d. case, the optimal set of energy thresholds can be found by a $N$-dimension exhaustive search from all the possible combinations of energy thresholds with the analytical expression derived in (\ref{outage1}). The computation complexity is $O\left(L^N\right)$, which grows exponentially with the number of relays $N$. Thus, finding the optimal set of energy thresholds becomes computationally prohibitive when the number of relays $N$ and the level of batteries $L$ are large. To overcome this problem, in the following subsection, we provide a heuristic approach to design the energy thresholds for the general i.n.i.d. case.

\subsection{A Heuristic Approach for i.n.i.d Scenario}
Intuitively, a higher energy threshold and forwarding transmit power should be set for those relays closer to the source node. This is because that the average amount of harvested energy is higher for those relays compared to the ones far away from the source node. On the other hand, relays closer to the source are relatively further away from the destination node such that the associated second hop channels are relatively weaker, a larger transmit power should be used to overcome the higher path loss. On the other hand, for the relays near to the destination node, smaller forwarding transmit power can be adopted due to their limited harvested energy and stronger second hop channels. Inspired by this fact, we set the energy consumption for information forwarding at each relay $R_u$ as
\begin{equation}
\tilde \beta_u = z \lambda_{S{R_u}} / \lambda_{{R_u}D},
\end{equation}
where $z$ is a scalar factor to adjust the overall transmit power of all the relays. For the considered discrete-battery model, the designed energy consumption for information forwarding $\tilde \beta_u$ should be discretized to one specific energy level of the battery. Note that the definition of discretization given below (\ref{dif}) no longer holds as we cannot simply choose $\beta_u \in \left\{{\varepsilon _1}, {\varepsilon _2}, \cdots, {\varepsilon _L}-\alpha \right\}$. We now define the discretized value of $\tilde \beta_u$ for the $u$-th relay as ${\beta_u} \buildrel \Delta \over = {\varepsilon _{j}}$,
\begin{equation}
j=\left\{
\begin{matrix}
\begin{split}
&\arg \mathop {\min }\limits_{i \in \left\{ {0,1, \cdots ,L} \right\}} \bigg\{{\varepsilon _i}:{\varepsilon _i} \ge \tilde {\beta_u}\bigg\},  \text{if} \quad \tilde {\beta_u} \le {\varepsilon _L}-\alpha \\
&L-{\alpha  \over {\varepsilon _1}} , \text{if} \quad \tilde {\beta_u} > {\varepsilon _L}-\alpha
\end{split}
\end{matrix}
\right..
\end{equation}
Based on the above definitions, the discretized energy threshold for relay $R_u$ can be further expressed as
\begin{equation}\label{threshold}
{\chi_u}= \alpha+\beta_u = \alpha+ \min \left( {{\varepsilon _1}\left\lceil {z  \lambda_{S{R_u}} \over {\lambda_{R_u D} {\varepsilon _1}}} \right\rceil ,C-\alpha} \right).
\end{equation}

Similar to the analysis given in the previous subsection, we can deduce that there should exist an optimal value of $z$ that minimizes the system outage probability. To find the optimal value of $z$,  we first define $\lambda_{\max} $ and $\lambda_{\min}$ as the maximum and minimum value of the term $\lambda_{S{R_u}} / \lambda_{{R_u}D}$, among all relays, respectively. The optimal $z$ should exist within the interval $\left[{{\varepsilon _1}/\lambda_{\max}  , {{\varepsilon _L}/\lambda_{\min} } }\right]$, where $z = {\varepsilon _1}/\lambda_{\max}$ could make all relays choose $\varepsilon_1 +\alpha$ as their energy thresholds and $ z = {{\varepsilon _L}/\lambda_{\min} }$ will force all relays to adopt energy level $L$ as their energy thresholds. The optimal $z$ thus can now be achieved by performing a one-dimension exhaustive search from this interval over the derived outage probability expression. In order to capture all the possible combinations, we search $z$ with an increment of ${\varepsilon _1}/\lambda_{\max}$ each time and the computation complexity for the proposed heuristic approach can be expressed by $O\left(L\lambda_{\max}/\lambda_{\min}\right)$. Note that the complexity of the heuristic approach is obviously much lower than that of the exhaustive search given by $O\left(L^N\right)$.

\section{Performance Upper Bound}\label{sec:bounds}
As described in Section \ref{sec:MC}, we adopt a practical finite-capacity battery model in this paper. With this model, it can be readily deduced that the system performance could be improved when we increase the battery capacity (i.e., $C$). This is because a larger battery capacity can reduce the energy loss caused by energy overflow (i.e., the battery cannot be charged when it is full) and thus the relays have more available energy to support their information forwarding operation. On the other hand, we can infer that the system performance improvement speed actually decreases as the battery capacity increases since the energy overflow happens more rarely when the battery capacity keeps increasing. A natural question that comes up here is ``\emph{For a given network setup, how large of the battery capacity $C$ will be sufficient?}" This question is particularly important for the considered system as one of its potential applications will be low-cost and lower-power networks (e.g., wireless sensor networks), in which the network deployment cost should be kept as low as possible by carefully selecting the battery capacity. However, it is hard to find the answer of this question based on the derived expressions in the previous section due to their complexity. We are thus motivated to adopt an indirect way: We first derive performance upper bounds of the considered system and the \emph{sufficient} battery capacity will then be obtained when a certain value of $C$ can make the performance expressions derived in previous sections approach their corresponding upper bounds. In this sense, in this section we analyze the performance upper bounds of the considered system with infinite battery capacity (i.e., $C \to \infty$ and implicitly $L  \to \infty$).


When the relay batteries are infinite, there will be no energy overflow. As such, we can implement the well-known flow conservation law to evaluate the system outage probability. Specifically, in our ETMRS scheme, the relay battery is charged if operated in EH mode and discharged if operated in IF mode. In this sense, for each relay, the total amount of harvested energy in a long term should equal to the total amount of energy consumed for information decoding and forwarding. Mathematically, we have the following formula for the $u$-th relay
\begin{equation}\label{supercondition}
{p_u} \bar {E}_u = \left(1-p_u\right)\left[\left(1-q_u\right)\alpha + q_u \left(\alpha+{\beta_u}\right)\right],
\end{equation}
where $p_u$ denotes the probability that relay $R_u$ performs EH operation, $q_u$ denotes the probability that the $u$-th relay falls in the decoding set, $\bar {E}_u$ is the average amount of harvested energy during each EH operation, $\alpha$ is the circuit energy consumption for each decoding operation and ${\beta_u}$ is the energy consumption for information forwarding of $R_u$ adopted in the finite-capacity case. 
Since both $C$ and $L$ are infinity, the term $\bar {E}_u$ can be obtained from (\ref{EH}) given by
\begin{equation}\label{eu}
\bar {E}_u = {1 \over 2}\eta P\lambda_{SR_u}.
\end{equation}

With the fact that relay $R_u$ falls in the decoding set only when it performs IF operation and decodes the source's information correctly, we thus have
\begin{equation}\label{superrelation}
q_u= \left({1-p_u}\right) \left({{1-\Pr \left\{{\varphi_u = 0}\right\}}}\right),
\end{equation}
where the term $\Pr \left\{{\varphi_u = 0}\right\}$ has been derived in (\ref{general}). By jointly considering (\ref{supercondition})-(\ref{superrelation}), we can now obtain $q_u$ expressed as
\begin{equation}\label{q_u}
q_u = {1 \over {{1 \over {1-\Pr \left\{{\varphi_u = 0}\right\}}}+ {{2\alpha  + 2{\beta _u}\left( {1 - \Pr \left\{ {{\varphi _u} = 0} \right\}} \right)} \over {\eta P{\lambda _{S{R_u}}}\left( {1 - \Pr \left\{ {{\varphi _u} = 0} \right\}} \right)}}}}.
\end{equation}

The first and second probability terms in the system outage probability defined in (\ref{outage1}) can now be evaluated by $q_u$ as
\begin{equation}\label{firsttermasy}
\begin{split}
\Pr \left\{ \emptyset   \right\} = \prod\limits_{u:R_u  \in \Lambda }  \left({1-q_u}\right),
\end{split}
\end{equation}
\begin{equation}\label{secondtermasy}
\begin{split}
\Pr \left\{ {{\Phi _{k,n}}} \right\} = \prod\limits_{u:R_u  \in \Phi_{k,n} } q_u  \prod\limits_{u:R_u  \notin \Phi_{k,n} }  \left({1-q_u}\right).   \\
\end{split}
\end{equation}

The third probability term in (\ref{outage1}) is independent of $C$ and $L$, and only depends on the energy threshold of the relays. In this sense, the third probability term in (\ref{outage1}) remains the same as the finite-capacity case, which has been derived in (\ref{2}). We thus can express the upper bound of system outage probability for the proposed ETMRS scheme corresponding to infinite battery capacity as (\ref{asymout}) on top of the next page.
\begin{figure*}[!t]
\begin{equation}\label{asymout}
\begin{split}
{P_{\rm{out}}^{ub}} \approx  \prod\limits_{u:R_u  \in \Lambda } \left({1-q_u}\right) + \sum\limits_{k = 1}^N {\sum\limits_{n = 1}^{\binom{N}{k}} {\left[{\prod\limits_{u:R_u  \in \Phi_{k,n} } q_u  \prod\limits_{u:R_u  \notin \Phi_{k,n} } \left({1-q_u}\right)}\right] }} \left[{1 - \exp ( - a_{k,n} v) \sum\limits_{i = 0}^{k - 1} {{\left({a_{k,n} v}\right)^{i} } \over {i!}}}\right].
\end{split}
\end{equation}

\hrulefill
\vspace*{4pt}
\end{figure*}

When the special i.i.d. channel fading case is considered, similar to the above analysis, the system outage probability is upper bounded by (\ref{asymoutiid}) on top of the next page.
\begin{figure*}[!t]
\begin{equation}\label{asymoutiid}
\begin{split}
{\mathord{\buildrel{\lower3pt\hbox{$\scriptscriptstyle\frown$}}\over
 P} _{\rm{out}}^{ub}} \approx \left({1-q}\right)^N  + \sum\limits_{k = 1}^N { {\binom{N}{k}} q^k \left({1-q}\right)^{N-k} \left[{1 - \exp ( - {{{{{{N_0 v}} \over { 4 k \beta \sigma^2} }}} }) \sum\limits_{i = 0}^{k - 1} {{\left({{{{{{{N_0  v}} \over { 4 k \beta \sigma^2} }}} }}\right)^{i} } \over {i!}}}\right]},
\end{split}
\end{equation}

\hrulefill
\vspace*{4pt}
\end{figure*}

where $q$ is the probability of each relay falling in the decoding set when the channels are i.i.d. and it can be easily derived based on (\ref{q_u}) and expressed as
\begin{equation}\label{q}
q = {1 \over {{1 \over {1-\Pr \left\{{\varphi = 0}\right\}}}+  {{2\alpha  + 2{\beta}\left( {1 - \Pr \left\{ {{\varphi} = 0} \right\}} \right)} \over {\eta P{\lambda _{S{R}}}\left( {1 - \Pr \left\{ {{\varphi} = 0} \right\}} \right)}}}}.
\end{equation}

\begin{remark}
By substituting (\ref{general}) into (\ref{q_u}), we can see that the probability for the $u$-th relay falling in the decoding set (i.e., $q_u$) is proportional to the term $P\lambda_{SR_u}$. This is understandable since as the value of $P\lambda_{SR_u}$ increases, the $u$-th relay harvests more energy on average and is more likely to decode the source signal successfully in the first hop. Moreover, as expected, the value of $q_u$ is inversely proportional to that of the energy threshold $\beta_u$. Furthermore, using the above performance upper bounds of system outage probability, we are able to judge whether a certain value of battery capacity $C$ is sufficiently large for a given network setup via numerical results as shown in next section.

\end{remark}


\section{Simulations and Discussions}
In this section, we present some simulation and numerical results to validate and illustrate the above theoretical analysis. In order to capture the effect of path-loss, we use the model that ${\lambda _{XY}} = {10^{-3} \over {1 + {d_{XY}^\omega}}}$, where ${\lambda _{XY}}$ is the average channel power gain between node $X$ and $Y$, ${d_{XY}}$ denotes the distance between node $X$ and $Y$, and $\omega  \in \left[ {2,5} \right]$ is the path-loss factor. Note that a 30 dB average signal power attenuation is assumed at a reference distance of 1 meter (m) in the above channel model \cite{hechen}. For simplicity, we consider a linear topology that the relays are located on a straight line between the source and destination and denote by $\pmb{d_{SR}} = \left\{ {d_{SR_1 } ,d_{SR_2 } , \cdots, d_{SR_N } } \right\}$ the set of distances between source and all relays. We use $\pmb{\chi} = \left\{ \chi_1 , \chi_2 , \cdots, \chi_N  \right\}$ to represent the energy threshold set for all the relays. In all the following simulations, we set the distance between the source and destination ${d_{SD}} = 20$m, the path-loss factor $\omega  = 3$, the severity parameter $m_u=2$, $\forall u$, the noise power ${N_0} =  - 90$dBm, and the energy conversion efficiency $\eta  = 0.5$.

\begin{figure}
\centering
  {\scalebox{0.40}{\includegraphics {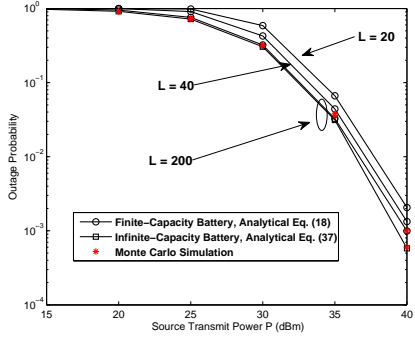}
  \label{fig:montsimu1}}}
\caption{The outage probability of the proposed ETMRS scheme versus the source transmit power for different battery levels $L$, where transmission bit rate $\kappa=1$, $C=2 \times 10^{-5}$, $\alpha = 10^{-7}$, $\pmb {d_{SR}} = \left\{5,5.5,6,6,6,6,6.5,7\right\}$ and $\pmb {\chi} = \left\{3,3, 3,3,3,3,4,4\right\} \times 10^{-6}$.
\label{montsimu}}
\end{figure}

We first compare the analytical system outage probability with its Monte Carlo simulation, which corresponds to the case that the charging of the relay batteries is continuous (i.e., $L \to \infty$). To this end, we plot the system outage probability of our ETMRS scheme versus the source transmit power for different battery levels $L$ in Fig. \ref{montsimu}. We can see that the derived analytical expression of outage probability approaches the corresponding Monte Carlo simulation result as the discrete battery level $L$ increases. Specifically, when $L=200$, the analytical expression coincides well with the simulation result, which verifies the effectiveness of the adopted MC model and the correctness of our theoretical analysis presented in Sec. II-IV. Moreover, as expected, the performance of the ETMRS scheme with finite-capacity battery is bounded by the one with infinite-capacity battery. As the analytical results agree well with the simulation results and for the purpose of simplicity, in the following, we will only plot the analytical results of the ETMRS scheme.

\begin{figure}
\centering
 \subfigure[Combinations of $P$ and $d_{SR_u}$]
  {\scalebox{0.40}{\includegraphics {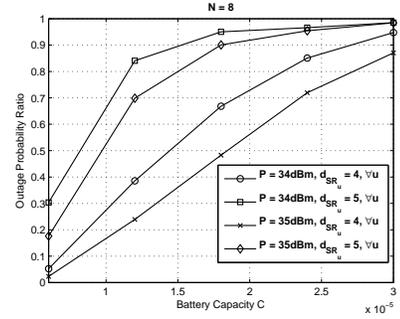}
  \label{fig:csimu1}}}
\hfil
 \subfigure[Combinations of $N$ and $d_{SR_u}$]
  {\scalebox{0.40}{\includegraphics {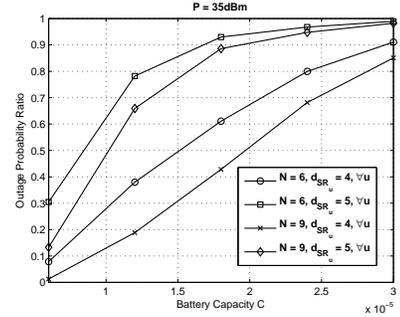}
\label{fig:csimu2}}}
\caption{The ratio between the upper bound outage probability of the ETMRS scheme with infinite battery capacity and the outage probability with finite battery capacity versus battery capacity $C$ for different source transmit power and relay topologies, where $\kappa = 1$, $\alpha = 10^{-7}$, $\chi_u = 4 \times 10^{-6}$, $\forall u$ and $L=600$.
\label{csimu}}
\end{figure}

Recall that we derived system performance upper bound in Section \ref{sec:bounds}, which can be used to judge whether a certain value of $C$ is sufficiently large via numerical results. To show this, we now plot the outage probability ratio of the proposed ETMRS scheme with finite battery capacity over its upper bound with infinite battery capacity versus relay battery capacity $C$ in Fig. \ref{csimu}. For simplicity, we consider the special i.i.d. case where the relays are located in a cluster. The outage probability ratio is formally defined as ${\mathord{\buildrel{\lower3pt\hbox{$\scriptscriptstyle\frown$}}\over
 P} _{\rm{out}}^{ub}} / {\mathord{\buildrel{\lower3pt\hbox{$\scriptscriptstyle\frown$}}\over
 P} _{\rm{out}}} \in (0,1]$, where ${\mathord{\buildrel{\lower3pt\hbox{$\scriptscriptstyle\frown$}}\over
 P} _{\rm{out}}^{ub}}$ derived in (\ref{asymoutiid}) is the system outage probability for infinite-capacity battery case and ${\mathord{\buildrel{\lower3pt\hbox{$\scriptscriptstyle\frown$}}\over
 P} _{\rm{out}}}$ given in (\ref{iid1}) is the system outage probability for finite-capacity battery case. From Fig. \ref{csimu}, we can first observe that the outage probability ratio monotonically increases as the battery capacity $C$ grows and gradually converges to 1 when the value of $C$ is large enough. This indicates that the performance gap between the finite-capacity battery and infinite-capacity battery decays to 0 as $C$ increases. However, the convergence speed varies for different network setups. Specifically, from Fig. \ref{fig:csimu1}, we can see that the outage probability ratio converges to 1 slower with either higher source transmit power or shorter distances between source and relays. This can be understood as follows. The amount of harvested energy at relays increases when the source transmit power increases or the distances between source and relays reduce. In this case, a larger battery capacity is required to avoid the potential energy overflow (i.e., the battery cannot be charged when it is full). From Fig. \ref{fig:csimu2}, we find out that as the number of relays $N$ increases, the convergence speed is also reduced. This is understandable since energy overflow happens in a higher probability as $N$ grows. Similarly, larger value of $C$ is required to make the outage probability ratio close to 1. We can now summarize that larger capacity batteries should be equipped at energy harvesting relays for those network setups with higher source transmit power, shorter distances between source and relays or larger number of relays.

\begin{figure}
\centering
 \subfigure[Combinations of $P$ and $d_{SR_u}$]
  {\scalebox{0.40}{\includegraphics {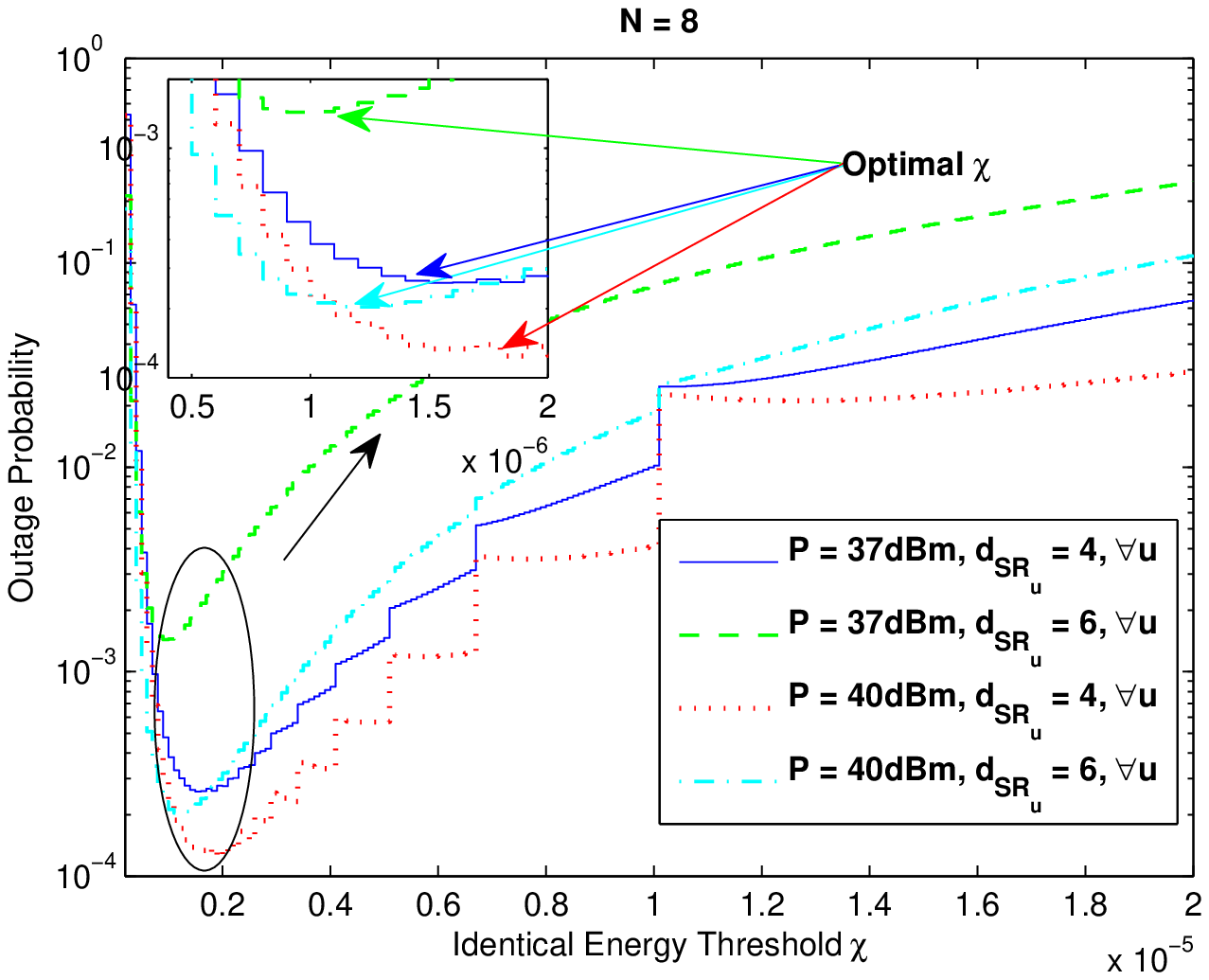}
  \label{fig:tplot1}}}
\hfil
 \subfigure[Combinations of $N$ and $d_{SR_u}$]
  {\scalebox{0.40}{\includegraphics {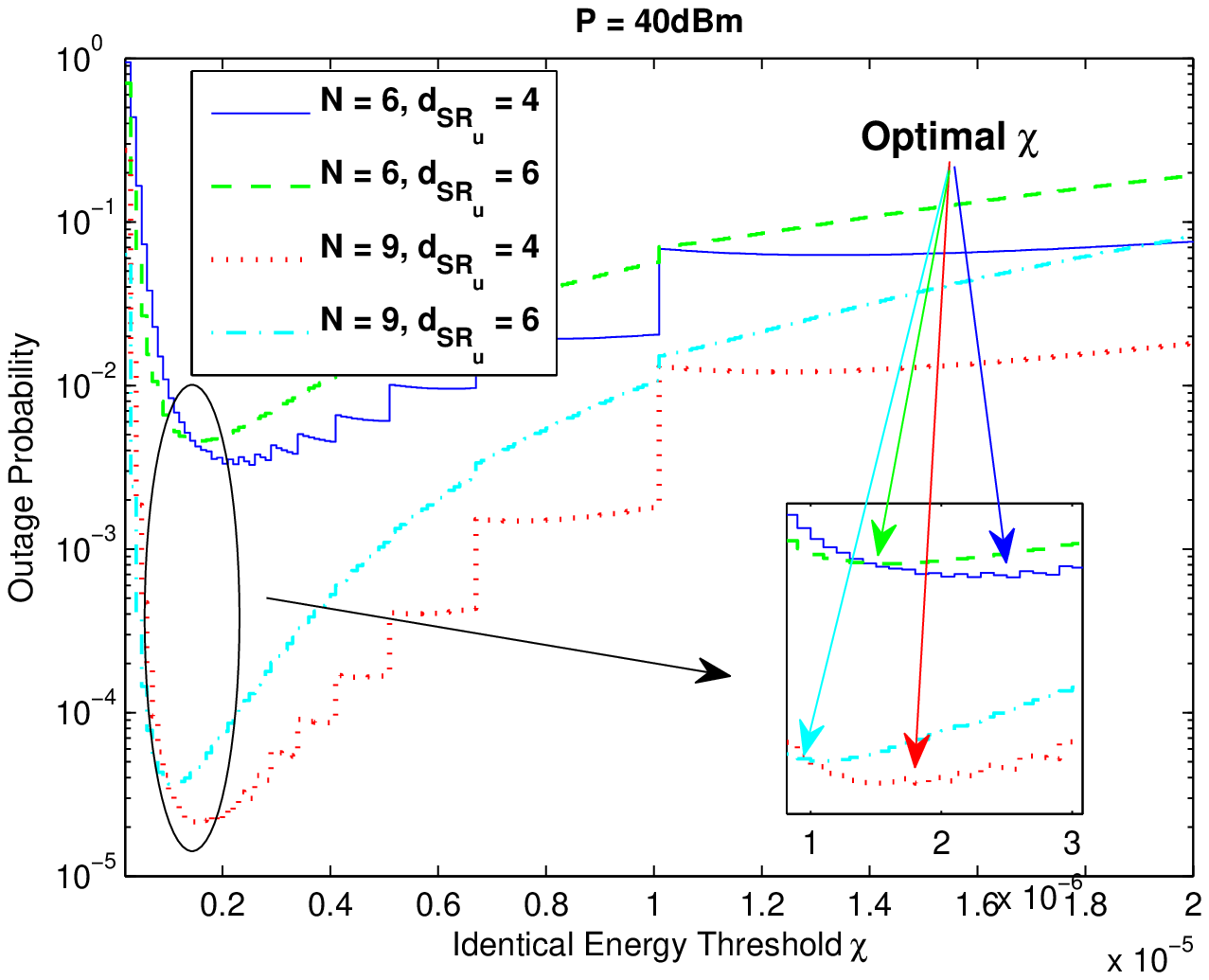}
\label{fig:tplot2}}}
\caption{The outage probability of the proposed ETMRS scheme versus the identical energy threshold for different source transmit power, relay topologies where $\kappa=1$, $C=2 \times 10^{-5}$, $\alpha = 10^{-7}$ and $L=200$.
\label{tplot}}
\end{figure}

In Fig. \ref{tplot}, we plot the outage probability versus the energy threshold of the i.i.d. channel fading case for different source transmit power and relay location. Recall that the energy threshold $\chi$ is discretized to one of the $L+1$ energy levels of the battery excluding the empty level. Thus, the outage probability is plotted in stair curve in the figure. First of all, Fig. \ref{tplot} demonstrates that there exists an optimal energy threshold that minimizes the outage probability in all considered cases, which validates our deduction in Remark 1. Moreover, we can see from Fig. \ref{fig:tplot1} that the higher the transmit power at the source, the larger the value of optimal energy threshold. This is because the relays can harvest more energy when the transmit power of the source increases and thus a larger energy threshold can be supported. We can also see from Fig. \ref{fig:tplot1} that a smaller energy threshold should be chosen when the relays are located away from the source. This can be explained as two folds. Firstly, the harvested energy at the relays is limited when they are far away from the source. Secondly, the second hop channel condition becomes better when the relays are away from the source (i.e., close to the destination) and a small energy threshold is enough to avoid system outage. The number of the relays $N$ also affects the optimal energy threshold. From Fig. \ref{fig:tplot2} we can see that as the number $N$ increases, the optimal energy threshold decreases. This is understandable since in cooperative relay networks, a decoding set with more relays can achieve the same outage probability with less transmit power. Note that similar results can be observed for the general i.n.i.d case, which are not provided here due to space limitation..
\begin{figure}
\centering
  {\scalebox{0.40}{\includegraphics {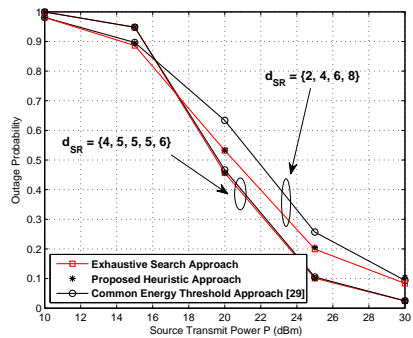}
  \label{fig:compare1}}}
\caption{Outage probability of the proposed ETMRS scheme and the common energy threshold scheme for different relay locations, where $\kappa = 1$, $C=2 \times 10^{-5}$, $\alpha = 10^{-7}$ and $L=20$.
\label{th}}
\end{figure}

We then compare the system performance of the proposed ETMRS scheme with optimized energy threshold and the common energy threshold scheme in Fig. \ref{th}. The optimal energy threshold of the ETMRS is obtained by the exhaustive search based approach and the proposed heuristic one. As the optimal energy threshold for the i.i.d. case can be easily obtained via a one-dimension exhaustive search, we only consider the general i.n.i.d. case where the optimal energy thresholds for the relays could be different. In Fig. \ref{th}, we plot the system outage probability for these two approaches versus the source transmit power for two different relay topologies. The outage probability of exhaustive search approach is obtained via a $N$-dimension exhaustive search from all the possible combinations of energy thresholds. Due to the intense computation complexity mentioned before, we consider the case that each relay only has 20 energy levels. For the proposed heuristic approach, the computation complexity can be dramatically reduced and it is thus particularly suitable to those networks with large number of relays and energy levels. From Fig. \ref{th}, we can observe that the proposed heuristic approach with reduced complexity can achieve the near optimal system performance in both simulated scenarios. We also can see that our proposed scheme outperforms the common-energy strategy for all simulated cases. For the case that the relays are located closely, our proposed scheme and the one in \cite{five-multi-relay} performs similarly. When the relays are located differently, our proposed scheme improves the performance significantly by using different energy thresholds for each relay.

\begin{figure}
\centering
  {\scalebox{0.40}{\includegraphics {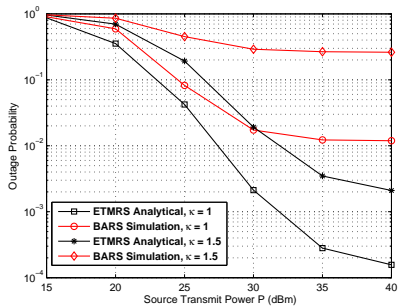}
  \label{fig:compare1}}}
\caption{The outage probability of the proposed ETMRS scheme and the existing BARS scheme with optimal settings versus source transmit power for different transmission bit rate, where $N=8$, $d_{SR_u}=3,~\forall u$, $C=2 \times 10^{-5}$, $\alpha = 10^{-7}$ and $L=200$.
\label{compare}}
\end{figure}

Finally, we compare the proposed ETMRS scheme with the existing BARS scheme \cite{or-eaccumu} in Fig. \ref{compare}. As the BARS scheme proposed in \cite{or-eaccumu} only considers a clustered topology, we thus compared BARS scheme with the special i.i.d. case of the proposed ETMRS scheme. In Fig. \ref{compare}, we plot the optimal outage probabilities of the ETMRS and BARS schemes versus source transmit power for different transmission rate. It can be observed that the proposed ETMRS scheme can achieve a lower outage probability than the BARS scheme. And their performance gap is enlarged as the transmission rate grows.


%
\section{Conclusions}
In this paper, we proposed an energy threshold based multi-relay selection (ETMRS) scheme for accumulate-then-forward energy harvesting relay networks. We modeled the finite-capacity battery of the relays by a finite-state Markov Chain (MC) in order to evaluate their stationary  distribution. We then derived an approximate analytical expression for system outage probability of the proposed ETMRS scheme over independent but not necessarily identical mixed Nakagami-$m$ and Rayleigh fading channels. A heuristic approach was then designed to minimize the system outage probability, which was shown to achieve near-optimal performance with reduced computational complexity. Moreover, we derived upper bounds for the concerned system performance corresponding to the case that all relays are equipped with infinite-capacity batteries. Numerical simulations validated the accuracy of the analytical results, demonstrated the impact of various system parameters and provided some insights of practical relay battery design. Numerical results showed that larger capacity battery should be equipped at energy harvesting relays for those network setups with higher transmit power, shorter distance in the first hop or larger number of relays. Furthermore, the proposed ETMRS scheme can considerably outperform the existing single relay selection scheme and the common energy threshold scheme.
\appendices
\ifCLASSOPTIONcaptionsoff
  \newpage
\fi

\bibliographystyle{IEEEtran}
\bibliography{References}
\end{document}